\documentclass{article}
\relax
\usepackage {graphics}
\begin{document}

\title{Influence Of Collapsing Matter On The Enveloping Expanding Universe\footnote{This work is dedicated to the memory of  Professor Albert Einstein the greatest scientist of our time in commemoration of the postulation of the special theory of relativity 100 years ago.} }         
\author{A. Latif  Choudhury\\
Department of Chemistry and Physics \\
Elizabeth City State University\\
Elizabeth City, NC 27909\\
 USA
}        
\date{April 27, 2005}          
\maketitle
\begin{abstract}
  {Using a collapsing matter model  at the center of an expanding universe as described by Weinberg we assume a special type of  generated   pressure. This pressure transmits into the surrounding expanding universe. Under certain restriction the ensuing  hubble parameter is positive. The deacceleration parameter fluctuates with time, indicating that the universe accelerates for certain time and decelerates for other time intervals.}
\end{abstract}   
\section{Introduction}       
  {$\;$$\;$In several earlier papers Choudhury[1,2,3,4] has constructed models where a Gidding-Strominger wormhole[5] is placed at the center of an expanding universe and showed that, under certain conditions, the physical universe can expand with calculable  acceleration. Following a different approach we now try to put a collapsing  matter cloud at the center of the physical universe. We assume that the collapsing cloud generates a pressure which obeys the classical adiabatic gas law. This pressure penetrates  the surrrounding universe. We add this pressure to the energy momentum tensor and compute the Hubble parameter and the deacceleration parameter, both of which are time dependent. The idea of the collapsing matter at the center of the universe is adopted from the treatment elaborated in the book of Weinberg [6].
We have shown in section 1 how the $R_c$, the scale factor of the core, is calculated. The parametric solution depends on a quantity $\psi$. In section 2 we have we have calculated the pressure by using the classical adiabatic gas law. In section 3 we are assuming the enveloping universe to be homogeneous and isotropic, ignoring the core of the collapsing matter. We obtained the scale factor ${R}(t)$. We first calculated the Hubble parameter and the deacceleration parameter without incorporating the pressure from  the core for large $t_0$ in section 4. We have shown that the enveloping universe is expanding with a constant deceleration. In section 5, we assumed that the pressure generated in the collapsing core is transmitted into the enveloping physical universe following classical Pascal's law. We adhoc added this pressure to the energy momentum tensor. We then recalculated both the Hubble  and the deacceleration parameter. Under certain restriction we have shown that the Hubble parameter can be positive and finite. The deacceleration parameter turns out to be fluctuating with time. In section 6 we discussed the significance of the result we obtained.}
\section{Pressure From  Collapsing Matter}
{$\;$$\;$We assume that a certain amount of matter of the density $\rho_c$ at the core of the physical universe is collapsing in accordance with Weinberg's treatment [6]. For the moment we ignore the pressure. In the comoving coordinate system the invariant interval is given by} 
\begin{equation}
d{{\tau}^{c2}}=-dt^2+{W^c}(r,t)d{r_c}^2+{V^c}(r,t)(d{\theta}_c^2+sin^2{\theta}_c d{\phi}_c^2).
\end{equation}
{With pressure negligible, we write }
\begin{equation}
T^{c\mu\nu}={\rho^c}U^{c\mu}U^{c\nu},
\end{equation}
{with}
\begin{equation}
U^{cr}=U^{c\theta}=U^{c\phi}=0, U^{ct}=1.
\end{equation}
{In the above equations the index "c" refers to collapsing entities. The Einstein's field equation then becomes}
\begin{equation}
R^c_{\mu\nu}=8{\pi}GS^c_{\mu\nu}. 
\end{equation}
{with}
\begin{equation}
S^c_{\mu\nu}= T^c_{\mu\nu}-{\frac1 2}g^c_{\mu\nu}{T^{c\lambda}}_\lambda. 
\end{equation}
{Following Weinberg we can show that}
\begin{equation}
W^c=R^{c2}(t)f^c(r)\; \;and \; \;V^c=R^{c2}(t)r^2.
\end{equation}
{with }
\begin{equation}
f^c(r)=(1-kr^2)^{-1}.
\end{equation}
{The equation $R^c(t)$ satisfies}
\begin{equation}
{\dot{R}}^{c2}(t)=\alpha[R^{c(-1)}(t)-1],
\end{equation}
{where}
\begin{equation}
\alpha={\frac {8{\pi}G} {3}}{\rho}^c(0).
\end{equation}
{The solution is the parametric equation of cycloid}
\begin{equation}
t=\frac{\psi+ sin{\psi}} {2\sqrt{\alpha}}.
\end{equation}
{and}
\begin{equation}
R^c={\frac 1 2}(1+ cos{\psi}).
\end{equation}
{We now introduce a simplified approximation assuming $\psi$ to be a small quantity. Then t takes the form}
\begin{equation}
t=\frac{\psi} {\sqrt{\alpha}},
\end{equation}
{and}
\begin{equation}
R^c=cos^2(\frac {{\sqrt{\alpha}}t} 2).
\end{equation}
{Before total collapse we assume that the core behaves as an adiabatic gas . The volume of the collapsing matter can be shown to be }
\begin{equation}
\tau^c=2{\pi^2}R^{c3}(t).
\end{equation}
{This thermodynamical core satisfies the adiabatic gas law}
\begin{equation}
P^c{\tau}^{c\gamma}= constant=B_1.
\end{equation}
{This yields}
\begin{equation}
P^c=Bcos^{-6\gamma}(\frac {\sqrt{\alpha}t} 2),
\end{equation}
{We conjecture that this pressure is transmitted into the surrounding expanding physical universe according to  Pascal's law.}
\section{Enveloping Space}
{$\;$$\;$We assume that the matter which surrounds the collapsing matter is homogeneous and isotropic. The extension of physical universe is so immense that if there is any violation of the properties of homogeneity and isotropy of the collapsing matter it can be ignored in our consideration. The space-time interval can be given by}
\begin{equation}
d{{\tau}^2}=-dt^2+R^2(t)[f(r)dr^2+ r^2d{\theta}^2+sin^2{\theta} d{\phi}^2).
\end{equation}
{The Einstein equation for the enveloping space would be}
\begin{equation}
R_{\mu\nu}=8{\pi}GS_{\mu\nu}, 
\end{equation}
{where}
\begin{equation}
S_{\mu\nu}= T_{\mu\nu}-{\frac1 2}g_{\mu\nu}{T^{\lambda}}_\lambda. 
\end{equation}
{with}
\begin{equation}
T_{\mu\nu}=Pg_{\mu\nu}+(P+{\rho})U_{\mu}U_{\nu}.
\end{equation}
{In Eq.(21) the U's are given by}
\begin{equation}
U^{r}=U^{\theta}=U^{\phi}=0, U^{t}=1.
\end{equation}
{The only nonvanishing compopnents of the Ricci tensor are}
\begin{equation}
R_{tt}=-\frac{3\ddot{R}} R, 
\end{equation}
\begin{equation}
R_{rr}={\frac {f'(r)} {r{f(r)}}}+{\ddot{R}}Rf(r)+2{\dot{R}}^2 f(r), 
\end{equation}
\begin{equation}
R_{\theta\theta}=1-{\frac1 {f(r)}}+{\frac{rf'(r)} {f^2(r)}}+{\ddot{R}}Rr^2+2{\dot{R}}^2r^2, 
\end{equation}
{and}
\begin{equation}
R_{\phi\phi}=R_{\theta\theta}sin^2(\theta). 
\end{equation}
{The righthand side of the components of Eq.(3.2) becomes}
\begin{equation}
S_{tt}={\frac{\gamma} 2}(\rho+3P), 
\end{equation}
\begin{equation}
S_{rr}={\frac{\gamma} 2}R^2f(-P+\rho), 
\end{equation}
\begin{equation}
S_{\theta\theta}={\frac{\gamma} 2}R^2r^2(-P+\rho), 
\end{equation}
{and}
\begin{equation}
S_{\phi\phi}={\frac{\gamma} 2}R^2r^2sin^2\theta(-P+\rho). 
\end{equation}
{In the above equations we have set}
\begin{equation}
{\gamma}=8{\pi}G. 
\end{equation}
{Taking the rr-component of the Einstein equation we get}
\begin{equation}
{\frac {f'(r)} {r{f(r)}}}+{\ddot{R}}Rf(r)+2{\dot{R}}^2 f(r)={\frac{\gamma} 2}R^2f(-P+\rho), 
\end{equation}
{Here we make a conjecture that the pressure and density of the physical universe only depends on time t. We find that the Eq.(32) yields}
\begin{equation}
2\sigma+{\ddot{R}}R+2{\dot{R}}^2 ={\frac{\gamma} 2}R^2(-P+\rho), 
\end{equation}
{where we define a constant $\sigma$ by the relation }
\begin{equation}
{\frac {f'(r)} {r{f^2(r)}}}=2\sigma. 
\end{equation}
{From $\theta\theta$-component setting}
\begin{equation}
2\beta={\frac1 r^2}-{\frac1 {r^2f(r)}}+{\frac{f'(r)} {rf^2(r)}}, 
\end{equation}
{we get}
\begin{equation}
2\beta+{\ddot{R}}R+2{\dot{R}}^2={\frac{\gamma} 2}R^2(-P+\rho). 
\end{equation}
{Subtracting from Eq.(33),  Eq.(36) we obtain}
\begin{equation}
2(\sigma-\beta)=2R^2\gamma P. 
\end{equation}
{The above equation yields}
\begin{equation}
P=\frac{\lambda} {{\gamma}R^2}, 
\end{equation}
{If we conjecture that $\sigma>\beta$, P will always be positive.}
\section{Hubble and Deacceleration Parameters Without The Influence Of A Collapsing Core}
{$\;$$\;$ In this section we start with the assumption that the collapsing core does exist, but the physical universe is expanding. To compute the Hubble parameter we turn to the tt-component of the Einstein Eq.(19). We get} 
\begin{equation}
\ddot{R}=-{\frac{\gamma} 6}R(\rho+3P). 
\end{equation}
{Combinig the above equation with Eq.(3.11)we get }
\begin{equation}
H(t)=\frac{\dot{R}} R=\sqrt{({\frac{\gamma\rho} 3}+2\gamma P-\frac{2\sigma} {R^2})}. 
\end{equation}
{Since the physical universe is, by assumption, expanding,we have for large $t\rightarrow {t_0}$, $R\rightarrow 0$.}
\begin{equation}
H(t_0) \rightarrow \sqrt{{\frac{\gamma\rho} 3}+2\gamma P}. 
\end{equation}
{For large $t_0$, the deacceleration parameter becomes:}
\begin{equation}
q_0(t_0) = -\frac{{\ddot{R}R}} {{\dot{R}}^2}= \frac{{\frac\gamma 6}(\rho+3P)} {\gamma(\frac\rho 3+2P)-\frac{2\sigma} {R^2}}. 
\end{equation}
{Since for large R, P vanishes, we obtain}
\begin{equation}
q_0 \rightarrow \frac1 2. 
\end{equation}
{Thus the universe is decelerating at a constant rate. But all cosmic observation indicates that at the present time the universe is expanding with acceleration [7]. Instead of introducing  the pressure from an expanding wormhole, we generate the pressure from a collapsing core as described in section 2. }
\section{Influence Of Pressure From Collapsing Matter}
{$\;$$\;$We now add the pressure which we conjectured in section 2, into the energy momentum tensor. This pressure originates from the core which is collapsing adiabatically. We take this pressure to be of the form $P_c$, derived in the Eq.(17). This pressure is transmitted into the enveloping universe by classical Pascal's law. We add this amount to the pressure in Eq.(40) and get}
\begin{equation}
H(t)=\frac{\dot{R}} R=\sqrt{{\frac{\gamma\rho} 3}+2\gamma (P+P_c)-\frac{2\sigma} {R^2}} 
\end{equation}
{Now from Eq.(3.16) $P \rightarrow 0$ as $R \rightarrow \infty$, we find for large time $t_0$}
\begin{equation}
H_0(t_0)=\sqrt{\gamma({\frac\rho 3}+2{B}cos^{-6\gamma}(\frac{\sqrt{{\alpha}{t_0}}} 2))}. 
\end{equation}
{We assume that the quantity ${\sqrt{{\alpha}{t_0}}} /2 $ is small even for large $t_0$ because $\alpha$ is very small. By choosing $\gamma$ appropriately we can make $H_0(t_0)$  fluctuate with time. There is a possibility that $H_0$ may become very large. In order to avoid such consequences we have to improve the solution of our approximation of Eqs.(11)and (12). With the addition of $P_c$ we find that the time dependent deacceleration parameter for an expanding universe becomes}
\begin{equation}
q_0(t_0)={\frac1 2}{ \frac{\rho{cos^{6\gamma}(\frac{{\sqrt{\alpha}}{t_0}} 2})+6B} {\rho{cos^{6\gamma}(\frac{{\sqrt{\alpha}}{t_0}} 2})+12B}} . 
\end{equation}
{If we choose $\gamma =1/6$, we can write }
\begin{equation}
q_0(t_0)={\frac1 2}[{1- \frac{6B} {\rho{cos(\frac{{\sqrt{\alpha}}{t_0}} 2})+12B_1}}] . 
\end{equation}
{This parameter is time dependent. The universe is expanding with acceleration if it satisfies the following condition:}
\begin{equation}
6B>{\rho{cos(\frac{{\sqrt{\alpha}}{t_0}} 2})+12B} 
\end{equation}
{In all other times the universe decelerates.} 
\section{Concluding Remarks}
{$\;$$\;$We have shown here that if we take the pressure and density of an expanding physical universe which are only time dependent, the Hubble time dependent parameter is positive. The deacceleration parameter for  such a universe turns out to be positive, indicating that the universe decelerates. As an alternative idea  we have put a gravitationally collapsing body at the center of the expanding universe and conjectured that the collapsing matter generates a pressure according to adiabatic gas law. This pressure transmits uniformly across the enveloping physical universe following Pascal's law of fluid pressure. For large $t_0$ satisfying the condition ${\sqrt\alpha }t_0$ to be small, the Hubble parameter stays positive and finite indicating expansion of the physical universe. The deacceleration parameter fluctuates with time. Under certain condition the universe accelerates. }
\section{References}
\begin{enumerate}
\item{A. L. Choudhury, Hadronic J.,23, 581 (2000).}
\item{L. Choudhury and H. Pendharkar, Hadronic J., 24, 275 (2001).}
\item{A. L. Choudhury, Hadronic J.,27, 387 (2004).}
\item{A. L. Choudhury: Wormhole core, extra dimensions,and physical universe; arXiv:gt-qc/0405135v1,27 May 2004.}
\item{S. B. Giddings and A. Strominger, Nucl. Phys. B307, 854 (1988).}
\item{S. Weinberg: Gravitation and Cosmology, J. Wiley and Sons, Inc., 342 (1972).}
\item{N. Bahcall, J. P. Ostriker, S. Perlmutter, and P. J. Steinhardt, Science, 284, 1481 (1999).} 
\end{enumerate}
\end{document}